Title

# Emergence of spin-orbit coupled ferromagnetic surface state derived from Zak phase in a nonmagnetic insulator FeSi

Authors

Yusuke Ohtsuka,[1]† Naoya Kanazawa,[1]*† Motoaki Hirayama,[1,2]† Akira Matsui,[1] Takuya Nomoto,[1] Ryotaro Arita,[1,2] Taro Nakajima,[2,3] Takayasu Hanashima,[4] Victor Ukleev,[5] Hiroyuki Aoki,[6,7] Masataka Mogi,[1]‡ Kohei Fujiwara,[8] Atsushi Tsukazaki,[8] Masakazu Ichikawa,[1] Masashi Kawasaki,[1,2] Yoshinori Tokura[1,2,9]

Affiliations

[1] Department of Applied Physics, The University of Tokyo, Tokyo 113-8656, Japan.
[2] RIKEN Center for Emergent Matter Science (CEMS), Wako 351-0198, Japan.
[3] Institute for Solid State Physics, The University of Tokyo, Kashiwa 277-8561, Japan.
[4] Neutron Science and Technology Center, CROSS, Tokai 319-1106, Japan.
[5] Laboratory for Neutron Scattering and Imaging (LNS), Paul Scherrer Institute (PSI), CH-5232 Villigen, Switzerland.
[6] Materials and Life Science Division, J-PARC Center, Japan Atomic Energy Agency, Tokai 319-1195, Japan.
[7] Institute of Materials Structure Science, High Energy Accelerator Research Organization, Tokai, 319-1106, Japan.
[8] Institute for Materials Research, Tohoku University, Sendai 980-8577, Japan.
[9] Tokyo College, University of Tokyo, Tokyo 113-8656, Japan.

*Corresponding author. Email: kanazawa@ap.t.u-tokyo.ac.jp
†These authors contributed equally to this work.
‡Present address: Department of Physics, Massachusetts Institute of Technology, Cambridge, MA 02139, USA

Abstract

A chiral compound FeSi is a nonmagnetic narrow-gap insulator, exhibiting peculiar charge and spin dynamics beyond a simple band-structure picture. Those unusual features have been attracting renewed attention from topological aspects. Although a signature of surface conduction was indicated according to size-dependent resistivity in bulk crystals, its existence and topological properties remain elusive. Here we demonstrate an inherent surface ferromagnetic-metal state of FeSi thin films and its strong spin-orbit-coupling (SOC) properties through multiple characterizations of the two-dimensional (2D) conductance, magnetization and spintronic functionality. Terminated covalent-bonding orbitals constitute the polar surface state with momentum-dependent spin textures due to Rashba-type spin splitting, as corroborated by unidirectional magnetoresistance measurements and first-principles calculations. As a consequence of the spin-momentum locking, non-equilibrium spin accumulation causes magnetization switching. These surface properties are closely related to the Zak phase of the bulk band topology. Our findings propose another route to explore noble-metal-free materials for SOC-based spin manipulation.

**Teaser**

Ferromagnetic-metal surface state derived from topological polarization and its spin-manipulation functionalities in FeSi.

**MAIN TEXT**

**Introduction**

Iron and silicon are earth-abundant elements, sustaining the modern industrial bases. Interesting magnetic and semiconducting properties of solid can be engineered through making Fe-Si based compounds, as exemplified by nearly half-metallic behavior of $Fe_3Si$ (*1*) and an optoelectronics-compatible direct-bandgap of *β*-$FeSi_2$ (*2*). A cubic chiral compound FeSi with the so-called *B*20-type structure has also attracted a great deal of interest for its unusual temperature (*T*) dependences of magnetic susceptibility (*3, 4*) and electrical conductivity (*5, 6*). While the density functional theory (DFT) predicts a non-magnetic and narrow-gap (~100 meV) insulating state (*7*), experimental results on magnetic susceptibility and conductivity spectrum indicate the presence of large effective moment (*3*) and the disappearance of charge gap above *T* = 200 K (*5*). Such spin and charge dynamics in FeSi cannot be explained by a simple thermal activation model based on the DFT band structure; this provokes interesting theoretical models incorporating strong correlation effects such as local Coulomb interactions (*8–11*) or formation of Kondo singlets (*5, 12*). The origins of bandgap and thermally-induced magnetism are still controversial partly due to strong sample dependence of physical properties (*6*).

Motivated by recent discoveries of topological aspects of related compounds (*vide infra*) which share similar magnetic and electronic properties as FeSi, we re-examine its unusual features by focusing on the surface state. One motive argument is about surface conduction in a Kondo insulator $SmB_6$ (*13*), which is attributed to the metallic surface state protected by SOC-assisted band inversion between 4*f* and 5*d* orbitals (*14*); thus $SmB_6$ is proposed to be a topological Kondo insulator (*15*). Another incentive is about Fermi-arc surface states observed in compounds with the FeSi-type crystal structure (*16–18*). The Fermi arcs arise to connect the surface projections of band-crossing points protected by the chiral crystalline symmetry (*19, 20*). In fact, previous experimental results on bulk FeSi hint at the existence of conducting surface state. While a number of photoemission spectroscopy (PES) studies reported a substantial density of states (DOS) at the Fermi level ($E_F$) (*21–24*), clear formation of gap structure (~60 meV) with much less residual DOS could be observed by UV-laser excited PES with the longer escape depth of the photoelectrons (~100 Å) and the better bulk-sensitivity (*25*). Such escape-depth-dependent PES spectra are indicative of the presence of a metallic surface producing a well-defined Fermi edge. Furthermore, a dramatic change in electrical resistivity with varying bulk sample size implied the metallic surface conduction, posing the possibility that FeSi would be a 3D topological insulating phase (*26*).

Unlike most studies using bulk crystals of submillimeter scale, we investigate FeSi epitaxial thin films with thicknesses less than several tens of nanometers to clarify the surface contributions. Through a systematic study on their thickness dependence and first-principles calculations on electronic and spin structures, we identify the surface ferromagnetic-metal state, which consists of covalent-bonding orbitals terminated at the surface, *i.e.*, dangling bonds. The terminated orbitals extend upward from the surface nuclei, resulting in potential gradient normal to the surface and large Rashba-type (*27*) surface band splitting (~35 meV). As consequences of the spin-momentum locking, current flow generates nonequilibrium spin density (*28*), which enables spintronic functionalities such as unidirectional magnetoresistance and efficient magnetization switching. The emergence of the spin-orbit-coupled surface state can be described in terms of the Zak phase (*29*), which is a central concept in the modern theory of electric polarization (*30–32*).

## Results

### Film-thickness dependence of electrical transport and magnetization properties

FeSi thin films with different thicknesses were grown epitaxially along the [111] direction on Si(111) substrates by using molecular beam epitaxy (MBE) method. We capped the films with 10-nm-thick MgO *in-situ* by radio-frequency magnetron sputtering to prevent oxidization or contamination in atmosphere. As-grown (uncapped) and Si-capped thin films were also prepared for comparing different interface conditions. (See Materials and Methods.) Unless otherwise noted, we discuss the results on MgO-capped FeSi thin films in this article.

We show the thickness ($t$) dependence of electrical transport and magnetization properties in Fig. 1. All the MBE films share common features: Resistivity $\rho_{xx}$ increases with decreasing temperature below $T = 300$ K (Fig. 1A), representing an insulating behavior; Hall conductivity ($\sigma_{xy}$) and magnetization ($M$) curves exhibit clear hysteresis loops with respect to magnetic-field ($H$) changes, indicating ferromagnetic ordering with out-of-plane anisotropy (Fig. 1, B and C). On the other hand, the magnitudes of these physical quantities are strongly dependent on $t$; this implies the heterogeneous electrical conduction and magnetization in the compositionally homogeneous thin films. (See fig. S1 for cross-sectional composition distribution.) Provided such heterogeneity stems from the coexistence of surface and bulk contributions, we can separate them by examining $t$-dependence of the measures per a sheet, *i.e.*, sheet conductance $\sigma_{xx}^{\text{sheet}} = t/\rho_{xx}$, sheet Hall conductance $\sigma_{xy}^{\text{sheet}} = \sigma_{xy}t$ and magnetization per surface unit cell area $M^{\text{sheet}}$. Given that the surface contribution is independent of $t$, it corresponds to the intercept of the sheet measure as a function of $t$. Meanwhile, the bulk contribution is in proportion to $t$, and hence accords with the slope. Figure 1, D–F display the $t$-dependence of $\sigma_{xx}^{\text{sheet}}$, $\sigma_{xy}^{\text{sheet}}$ and $M^{\text{sheet}}$. The $\sigma_{xx}^{\text{sheet}}$ vs $t$ curves have finite intercepts and show steeper slopes with increasing $T$ (Fig. 1D), which respectively represent surface conduction and an insulating behavior of bulk interior. The values of $\sigma_{xy}^{\text{sheet}}$ and $M^{\text{sheet}}$ at zero $H$ are almost constant despite the large variation of $t$ (Fig. 1, E and F). This represents that the ferromagnetic order is confined within the surface region and couples dominantly with the surface conduction, producing the *t*-independent anomalous Hall conductance as observed. Noticeably, the surface ferromagnetic order persists up to 200 K as seen in the inset of Fig. 1F, and the Hall angle $\sigma_{xy}^{\text{sheet}}/\sigma_{xx}^{\text{sheet}}$ of the surface anomalous Hall effect is ~0.014, relatively large in magnitude among conventional ferromagnetic metals (*33*).

**Observation of surface ferromagnetic state by polarized neutron reflectometry**

The surface ferromagnetic order was further verified by polarized neutron reflectometry (PNR) on a FeSi thin film with nominal thickness of *t* = 11 nm at *T* = 7 K in a magnetic field of 1 T applied parallel to the film surface. A polarized neutron beam obtained by a supermirror polarizer was reflected by the film in a specular geometry (Fig. 2A), and was detected by a two-dimensional position sensitive detector. The magnetic moments in the film were aligned parallel to the external magnetic field in the present condition, therefore we consider only specular and non-spin-flip scattering process in the following analysis. Polarization direction of the incident neutrons was set to be parallel or antiparallel to the external field at the sample position by a spin flipper and guide fields. We refer to the reflectivity measured with incident neutrons with up and down spins as $R^+$ and $R^-$, respectively. Figure 2B shows the curves of $R^+$ and $R^-$ as functions of the momentum transfer normal to the surface $Q_z$. A difference between $R^+$ and $R^-$, which stems from the depth profile of the finite net magnetization in the film, was clearly observed. By comparing the fidelity of candidate models assuming surface-, interface- or bulk-magnetization distributions (fig. S2), we identified that the magnetization is confined at the FeSi surface in contact with the MgO cap. The depth profiles of nuclear and magnetic scattering length density, which respectively represent the density distributions of the constituent elements and the magnetic moments, were estimated by fitting to PNR reflectivity curves with optimizing parameters (Fig. 2C, also see Materials and Methods). The fitting analysis indicates that the surface ferromagnetic layer has a thickness of 0.35 nm (with an error of +0.11 nm or −0.13 nm) and an average of 2.1 $\mu_B$/Fe (with an error of +0.4 $\mu_B$/Fe or −0.6 $\mu_B$/Fe), whose distribution is blurred by the structural roughness of 0.28 ± 0.02 nm, as schematically illustrated in the inset of Fig. 2A. This result is in good agreement with the calculated surface ferromagnetic state consisting mostly of the magnetic moments of the top three Fe layers (fig. S3).

***Ab initio* calculations on surface electronic and spin states**

To identify the origin of the ferromagnetic and conducting surface state, we performed *ab initio* calculations on electronic and spin states by employing the generalized gradient approximation (GGA) of the DFT. (See Materials and Methods for calculation setups.) In accord with the previous reports (*7*, *9–11*), the bulk band structure is of an insulating state with energy gap $\Delta$ ~ 80 meV (fig. S4). The valence bands contain strongly hybridized states between Fe 3*d* and Si 2*p* orbitals, whose Wannier functions are spread among Fe and Si atoms as exemplified in Fig. 3A, representing their covalent bonding character. On the other hand, the surface bands traversing $E_F$ are identified in the band structures of both paramagnetic and ferromagnetic FeSi slabs with a (111) plane (Fig. 3, C and E). The slab contains 15 repetitions of a stacking unit which is a quadruple layer (QL) consisting of Fe-dense, Si-dense, Fe-sparse and Si-sparse layers in the order from top to bottom (Fig. 3B) (*34*). As the number of surface states crossing $E_F$ between any pair of time reversal invariant momenta (TRIM) is even and their Fermi surfaces draw closed loops (Fig. 3D), they are not symmetry-protected surface states of a topological insulator or of a Weyl semimetal. The surface band is rather composed of dangling bonds associated with truncation of the covalent bonds; this is also verified by the disappearance of surface conductance and ferromagnetism in the Si-capped film, where the dangling bonds turn into the covalent bonds with Si (fig. S5). It is also remarkable that the ferromagnetic moment is generated only by the surface bands with exchange splitting (Fig. 3E and see again fig. S3).

As visualized in Fig. 3B, the eigenfunctions of surface states extend upward from the surface nuclei, forming an out-of-plane electric dipole moment. Such dipolar charge distribution induces large Rashba effect (*27*), resulting in spin-band splitting. Indeed, the Rashba-type spin splitting coexists with other types of SOC-induced spin splitting originating from lattice symmetry or exchange splitting due to the ferromagnetic order. In the paramagnetic state, the surface bands are split by up to 35 meV, which is surprisingly large in consideration of the small atomic numbers of Fe and Si (Fig. 3C). The Fermi contours are spin-polarized (Fig. 3D), resulting from a combination of different types of momentum-dependent spin polarizations: vortex-like in-plane spin texture due to the dominant Rashba effect (*27*), radial in-plane spin texture originating from the chiral-lattice symmetry (*35*), and threefold symmetric out-of-plane spin texture being reflective of (111)-surface geometry (*36*). In the ferromagnetic state, by contrast, the exchange splitting is dominant over the SOC splitting (Fig. 3E) and the above spin textures are perturbatively superposed on the collinear spin arrangement along *H*-direction (Fig. 3F).

**Detection of surface spin-momentum locking by unidirectional magnetoresistance measurements**

The characteristic spin-momentum locking is corroborated by *H*-direction dependence of unidirectional magnetoresistance (UMR) (*37–39*). The UMR is from an additional resistivity term $\Delta\rho_{xx}$ linear in electric current density $J$ and $H$; $\Delta\rho_{xx} = \rho_0 \gamma' J H$, $\rho_0$ and $\gamma'$ being the resistivity at zero $H$ and the coefficient, respectively. This term causes a nonreciprocal charge transport, where high and low resistance states are switched by the reversal of current direction (*37–39*). In a noncentrosymmetric system, where spin textures on Fermi surfaces have components odd in momentum $\boldsymbol{k}$, a charge current causes a shift of the Fermi surfaces by $\Delta\boldsymbol{k}$ along the current direction $x$ and hence induces a nonequilibrium overpopulation of spins in their polarized direction at the momentum $k_x$. As a consequence of the coupling between accumulated spins $\boldsymbol{S}(\Delta\boldsymbol{k})$ and $\boldsymbol{H}$, the resistance varies with the current direction $\hat{\boldsymbol{J}} \parallel \Delta\boldsymbol{k}$, which is phenomenologically expressed by $\Delta\rho_{xx}(\hat{\boldsymbol{J}}) \propto [\boldsymbol{S}(+\Delta\boldsymbol{k}) - \boldsymbol{S}(-\Delta\boldsymbol{k})] \cdot \boldsymbol{H}$. Owing to this nature, the momentum-dependent component of spin polarization can be determined by the *H*-angle dependence of UMR as demonstrated in SrTiO$_3$ (*36*), Bi$_2$Se$_3$ (*40*) and Ge (*41*).

We detect the UMR due to the spin-polarized Fermi contours (Fig. 3, D and F, schematic illustrations for Fig. 4, A and B) of FeSi thin film ($t = 5$ nm) in terms of second-harmonic resistance (*42*) $\rho_{xx}^{2f} = -\frac{1}{2}\rho_0\gamma'J_0H$ under $\mu_0|H| = 1$ T and an a.c. input current density $J = J_0 \sin 2\pi ft$ ($J_0 = 1.0 \times 10^8$ A/m$^2$, $f$ = 13 Hz, $\boldsymbol{J} \parallel [11\bar{2}]$). (See Materials and Methods.) Figure 4, C–E show the *H*-angle dependence of $\rho_{xx}^{2f}$ in the *yz*, *xy* and *zx* planes. The coordinates and the *H*-angle $\theta$ are defined in Fig. 4, C–E. In the paramagnetic state above *T* = 200 K, all the $\theta$-scans of $\rho_{xx}^{2f}$ follow sinusoidal functions, exhibiting their peak magnitudes at different angles: $\theta$ = 90°, 270° (*i.e.*, $\boldsymbol{H} \parallel \pm y$) for the *x-y* scans; $\theta$ = 0°, 180° (*i.e.*, $\boldsymbol{H} \parallel \pm z$) for the *z-x* scans; and $\theta$ = 90° + $\theta_0$, 270° + $\theta_0$ ($\theta_0$ = 12°) for the *y-z* scans. The net polarization of accumulated spins $\boldsymbol{S}(\Delta k_x)$ under $\boldsymbol{\hat{J}} \parallel x$ is therefore oriented in the direction tilted to the *z*-axis by $\theta_0$ from the *y*-axis, and its direction is reversed under the reversal of current direction. While the direction of $\boldsymbol{S}(\Delta k_x)$ in the *yz* plane represents the combination of the Rashba-type spin-momentum locking (*y* component) and the out-of-plane (*z*) component originating from the threefold symmetry of (111) plane, the parallel (*x*) component to electron momentum stemming from the chiral-lattice symmetry is not discerned as can be seen from the nearly zero $\rho_{xx}^{2f}$ at $\boldsymbol{H} \parallel x$ (Fig. 4, D and E). This may be because the films contain two crystalline domains with opposite chiralities (*43*), which leads to the opposite spin accumulations cancelling each other.

On the other hand, $\rho_{xx}^{2f}$ is not discernible in the ferromagnetic state below *T* = 200 K, regardless of *H*-direction. This is consistent with the expectation from the calculated spin texture (Figs. 3H and 4B), where the spins are polarized almost uniformly by the exchange interaction and the spin modulations, *i.e.*, the source of UMR in the paramagnetic state, are buried as perturbative components.

**Deterministic magnetization switching by electric current**

As one other consequence of the spin-momentum locking, we realize current-induced magnetization switching, which is mediated by the angular momentum transfer from the accumulated spins to magnetization torque, *i.e.*, spin-orbit torque (SOT) (*44*). Figure 5, A and B show the magnetization switching under small external magnetic fields $\mu_0H_x = \pm0.02$ T along the current direction *x* in the 5-nm-thick film. We employed current pulses $J_{\text{pulse}}$ for switching the perpendicular magnetization $M_z$ and evaluated the change in $M_z$ by Hall resistivity $\rho_{yx}$ measurement with a low current density after the current pulse injections. (See Materials and Methods.) Under $\mu_0H_x = 0.02$ T (Fig. 5B), the sign of $\rho_{yx}$ is switched from positive (negative) to negative (positive) at $J_{\text{pulse}} \sim 1.6 \times 10^{11}$ A/m$^2$ ($-1.6 \times 10^{11}$ A/m$^2$). The magnitude of $\rho_{yx}$ nearly reaches that in the fully polarized state, indicating the almost 100% switching ratio. The opposite $H_x$ results in the switching through opposite path (Fig. 5B). Given also that the switching does not occur under $\boldsymbol{H} \parallel y$ (fig. S6), a dampinglike SOT (*45*) $\tau_{\text{DL}} \propto \boldsymbol{M} \times (\boldsymbol{S_y} \times \boldsymbol{M})$ due to *y*-component of accumulated spins $\boldsymbol{S_y}$ is responsible for rotating $M_z$. Although $\boldsymbol{S_y}$ or the spin polarization due to Rashba effect is not discernible in the UMR measurement with $J \sim 1 \times 10^8$ A/m$^2$, its contribution emerges as the SOT in the high current density regime of $J \sim 1 \times 10^{11}$ A/m$^2$. Furthermore, the perpendicular magnetization can be repeatedly switched between up and down directions as shown in Fig. 5, C and D. It is also noteworthy that we observed the external-*H*-free magnetization switching in an as-grown thin film (figs. S7 and S8).

## Discussion

We have identified the metallic and ferromagnetic surface state and its spin-manipulation functionalities in the FeSi thin films. The surface state with the strong SOC stems from the truncation of the bulk orbitals spreading in the interatomic regions and the consequent appearance of surface polarization charges. The origin of the surface state can be interpreted in terms of the Zak phase (*29, 46, 47*), which is the well-defined quantity to represent the orbital position and plays a central role in the modern theory of electric polarization (*30–32*). A Zak phase is a Berry phase accumulated by a Bloch state along a path across the Brillouin zone. For describing surface polarization, the Zak phase is defined at each momentum $k_\parallel$ in the surface Brillouin zone by integrating Berry connection along the reciprocal lattice vector $G$ perpendicular to the surface: $\theta(k_\parallel) = -i\sum_n^{\text{occ}} \int_0^{|G|} \langle u_n(k) | \nabla_{k_\perp} | u_n(k) \rangle dk_\perp$, where $u_n(k)$ is the periodic part of the Bloch wavefunction in the *n*th band, with the gauge choice $u_n(k) = u_n(k+G)e^{iG\cdot r}$, and the sum is over the occupied states. As seen in the definition of surface polarization charge (*32*): $\sigma_{\text{surf}} = \boldsymbol{P} \cdot \boldsymbol{n} = \frac{e}{(2\pi)^3} \int_{\mathcal{A}} \theta(k_\parallel) dk_\parallel$ (modulo $e/A_{\text{surf}}$), the Zak phase corresponds to the displacement of the Wannier center, *i.e.*, the origin of bulk electric polarization $\boldsymbol{P}$, and also the resultant accumulation of surface charge $e\theta(k_\parallel)/2\pi$ (modulo $e$) at the given $k_\parallel$. Here, $A_{\text{surf}}$ and $\mathcal{A}$ are primitive surface cell areas in the real and reciprocal spaces; both are related by $A_{\text{surf}} = \mathcal{A}/(2\pi)^2$.

In Fig. 6, we simulate the Zak phase in the projected Brillouin zone by employing the face-centered cubic (fcc) lattice of Fe and Si, whose structure is obtained from the *B*20-type structure by a small shift of atomic positions (*7, 48*). Note that the size of the unit cell is 1/4 of the *B*20-type structure and the Zak phase is quantized to 0 or $\pi$ owing to the spatial inversion symmetry. The surface states appear near the Fermi level in the slab calculation (Fig. 6A), which is consistent with the Zak phase showing $\pi$ in most regions of the projected Brillouin zone (Fig. 6B). The existence of the surface state of FeSi is intrinsically associated with the Zak phase of its bulk band topology.

On applying this calculation to the actual case, the Zak phase becomes negligibly small in the entire Brillouin zone. This is because the present framework of the Zak-phase picture cannot cover the situation where an *even* number of the covalent bondings with $\theta(k_\parallel) = \pi$ are terminated in the surface unit cell. In such case, the calculated Zak phase yields $\sigma_{\text{surf}}(k_\parallel) = 2m \times e/2 = 0$ (modulo $e$) despite the actual surface charge accumulation. (*m* is an integer.) In the above simulation, we therefore reduce the Brillouin zone to half by slightly changing the atomic position so that there are an *odd* number of terminated orbitals. (It would be desirable to stretch the concept of Zak phase beyond the limitation of modulo $e$ arithmetic in the future.) Here we also note that no band inversion occurs in most region of the projected Brillouin zone upon the structural transformation from *B*20-type to fcc lattice (*48*) (also see fig. S9), which guarantees that the above model is sufficient for the qualitative discussion.

By reinvestigation based on the modern theory of electric polarization, a wide variety of materials will be identified to host functional surfaces of the Zak-phase origin. In contrast to the conventional idea that heavy elements are essential ingredients for SOC-based spintronic materials, the notion of Zak phase can shed light on compounds made of common elements and demonstrate another route to enhanced SOC utilizing surface states inherent in covalent crystals.

**Materials and Methods**

   **Thin-film growth and device fabrication**

   The FeSi thin films were grown on a highly resistive Si(111) substrate by the molecular beam epitaxy (MBE) method. For epitaxial growth of FeSi, we employed a FeSi(111) buffer layer of nominally 2 nm thickness, which was grown by reacting a deposited Fe layer with the Si(111)-(7×7) surface at 330 °C. We then co-evaporated Fe and Si onto this substrate below 330 °C, and followed this by an annealing at 350 °C. We deposited MgO-capping layer *in-situ* by radio-frequency magnetron sputtering at room temperature, while we evaporated Si by using the Knudsen cell for the fabrication of Si capping layer. Typical thickness of the capping layers was nominally 10 nm.

   $\theta$-$2\theta$ X-ray diffraction scans confirmed the epitaxial growth of $B$20-type FeSi. Cross-sectional transmission electron microscopy (TEM) images also verified the epitaxial growth of FeSi. Energy dispersive x-ray spectrometry (EDX) mappings showed the homogeneous distribution of Fe and Si in the thin films. See fig. S1 for the detailed characterization results.

   By using ultraviolet photolithography and Ar ion milling, we fabricated Hall bar devices with 1-mm width and 3-mm length for UMR measurements and ones with 10-μm width and 30-μm length for magnetization switching experiments. The electrodes Au (45 nm)/Ti (5 nm) were formed by electron beam deposition.

   **Magnetization and transport measurements**

   The magnetizations of rectangular samples (approximately 7.0 mm × 2.5 mm size) were measured by using the Reciprocating Sample Option (RSO) of a Magnetic Property Measurement System (MPMS, Quantum Design). The longitudinal and Hall resistivities were measured with a conventional four-terminal method by using the DC-transport option of a Physical Property Measurement System (PPMS, Quantum Design). The detailed temperature and magnetic-field dependences are shown in fig. S10.

   The unidirectional magnetoresistances were evaluated by measuring the second-harmonic signals of the a.c. resistances. We measured $\rho_{xx}^{2f}$ by using a current source (Model 6221, Keithley) and lock-in amplifiers (LI5650, NF Corporation). The current frequency was 13 Hz. The current amplitude was typically set to be 0.5 mA, which corresponds to a current density $1.0 \times 10^8$ A/m$^2$ in the 5-nm-thick film. The second harmonic voltages were anti-symmetrized as a function of *H*. The current-density dependence and the detailed temperature and magnetic field dependences of $\rho_{xx}^{2f}$ are shown in fig. S11.

   **Polarized neutron reflectometry**

   Polarized neutron reflectometry (PNR) was carried out at BL17 (SHARAKU) in the Materials and Life-science Experimental Facility of J-PARC in Japan (*49, 50*). The PNR measurements were performed on a rectangular sample of approximately 15 mm × 15 mm size at $T = 7$ K under an in-plane magnetic field $\mu_0 H_\parallel = 1$ T. Incident neutrons with a wavelength band of 2.2-8.8 Å were polarized either spin-up (+) or spin-down (−) with respect to the magnetic field. We measured the reflectivity for the incident neutrons with up or down spins ($R^+$ or $R^-$) without polarization analysis for the reflected neutrons. The spin-flip scatterings were not expected at 7 K and 1 T, because the magnetization of the sample was fully aligned along *H*. The depth profiles of nuclear and magnetic scattering length densities were deduced through model fitting to the PNR spectra by using GenX software (*51*), where layer thicknesses, densities, surface/interfacial roughness and magnetic moments were employed as the fitting parameters.

**Current-induced magnetization switching**

The current pulses with a 0.11-ms duration and varying pulse heights were injected for magnetization switching. The critical current to realize the magnetization switching ranged from 7 mA to 12 mA, which corresponds to current density of $1.4 \times 10^{11}$–$2.4 \times 10^{11}$ A/m$^2$ in the 5-nm-thick film. After each pulse injection, we monitored anomalous Hall resistivity with passing a low d.c. current of 0.05 mA, which corresponds to a much lower current density $1.0 \times 10^9$ A/m$^2$ in the 5-nm-thick film. Electric currents were supplied by a current source (Model 6221, Keithley) and voltages were detected by a voltmeter (Model 2182A, Keithley).

*Ab initio* **calculations**

The Wannier function visualizations shown in the main text are obtained from first-principles calculations. We first perform a calculation based on the DFT with Quantum Espresso code (*52*), where we assume a nonmagnetic case and adopt the unit cell of FeSi. The lattice parameters of FeSi are taken from the experiment and set $a = 4.448$ Å (*53*). We employ the exchange-correlation functional proposed by Perdew *et al.* (*54*) After completing the self-consistent calculation with the number of *k*-points $N = 12 \times 12 \times 12$ and the cutoff energy for the planewave basis set $E = 80.0$ eV, we perform the wannierization by using Wannier90 code (*55*). The inner and outer windows are set to [3,16.6] eV and [3,32] eV with respect to the Fermi level. The model consists of Fe-2*p*, Fe-3*d*, Si-2*s*, and Si-2*p* orbitals, which amounts to 48 Wannier functions in total for the unit cell. We extract the Wannier functions localized between the Fe and Si atoms and visualize it.

The band structure and the spin-texture visualization shown in the main text and the fig. S4 are obtained from first-principles calculations for slab systems. The actual calculations, with the SOC turned on, are performed based on DFT with VASP code (*56,57*), where we assume two different magnetic phases; paramagnetic and ferromagnetic. We take a unit cell of FeSi with the experimental lattice parameter, $a = 4.448$ Å (*53*), then consider its hexagonal extension, where the *c*-axis is taken parallel to the [111]-axis of the cubic cell.

The resulting structure consists of three sets of layers in the unit cell, each of which contains the layer with three Fe atoms, one Si atom, one Fe atom, and three Si atoms, and, therefore, twelve layers per unit cell in total. The supercell systems are constructed by stacking five unit-cells along the *c*-axis so that the surface layer consists of three Fe atoms per unit cell and three Si atoms on the opposite side. The overall structure extends to 61.6

Å, including a vacuum layer of 23.1 Å at the edge of the slabs. We employ the exchange-correlation functional proposed by Perdew *et al.* (*54*), $E_c$ = 400 eV as the cutoff energy for the plane-wave basis set, and $N = 6 \times 6 \times 1$ as the number of *k*-points for the self-consistent calculation. Using PyProcar (*58*), we perform the subsequent calculations, *i.e.*, band calculations along the high symmetry line with the denser *k*-mesh, and spin-texture calculations on the Fermi surface, while Vaspkit (*59*) is exploited to extract the Bloch wave function of the surface band at the M point and visualize it.

In the calculation of FeSi having the fcc structure, we fix the lattice constant and change the coordinate of each atom. We employ $N = 12 \times 12 \times 12$ *k*-points for the bulk calculation and $N = 12 \times 12 \times 1$ *k*-points for the slab calculation. In the bulk calculation, Fe is located at the center of the unit cell. The supercell systems are constructed by stacking fifteen unit-cells.

**Acknowledgments**

We thank G. Aeppli, H. M Rønnow, T. Ideue, K. Kondou and T. Yokouchi for fruitful discussions. The PNR measurements at the Materials and Life Science Experimental Facility of the J-PARC was performed under a user program (Proposal No. 2020A0211).

**Funding:** This work was supported by JSPS KAKENHI (Grants No. JP20H01859, JP20H05155, JP20H01867), JST CREST (Grant No. JPMJCR16F1 and No. JPMJCR1874).

**Author contributions:** Y.O. and N.K. grew the thin films with support from K.F., A.T., M.I. and M.K. Y.O. and N.K. performed magnetization and transport measurements. Y.O., N.K., T.N., and T.H. performed polarized neutron reflectometry with support from V.U. and H.A. Y.O., N.K. and M.M. performed magnetization switching experiments. M.H., A.M, T.N. and R.A performed first-principles calculations. N.K., M.H. and Y.T. wrote the draft. N.K. and Y.T. conceived the project. All the authors discussed the results and commented on the manuscript.

**Competing interests:** Authors declare that they have no competing interests.

**Data and materials availability:** All data are available in the main text or the supplementary materials. The data may be available from the authors upon reasonable request.

# Figures and Tables

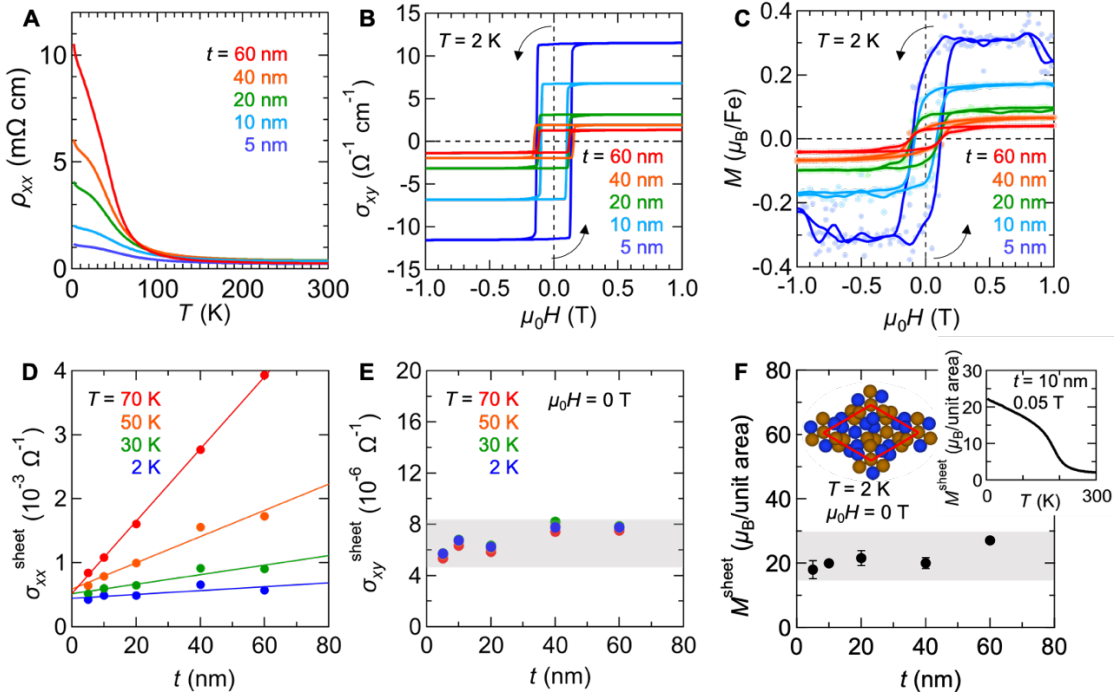

**Fig. 1. Thickness (*t*) dependence of transport and magnetization properties.** (A–C) Temperature (*T*) dependence of resistivity $\rho_{xx}$ at zero magnetic field (A), and magnetic-field (*H*) dependences of Hall conductivity $\sigma_{xy}$ (B) and magnetization *M* (C) at *T* = 2 K in FeSi thin films with various *t*. Here, the physical quantities per volume ($\rho_{xx}$, $\sigma_{xy}$ and *M*) are estimated under an assumption of the uniform electrical conductance and magnetization across the samples. The large *t*-dependences of their magnitudes indicate the heterogeneity of transport and magnetization properties. (**D–F**) Thickness dependences of sheet conductance $\sigma_{xx}^{\text{sheet}}$ (D), sheet anomalous Hall conductance $\sigma_{xy}^{\text{sheet}}$ (E) and remanent magnetization per surface unit cell area $M^{\text{sheet}}$ (F) at various *T*. Color solid lines or thick gray lines in each panel are guides to the eye. The insets of panel (F) show the surface unit cell and *T*-dependence of $M^{\text{sheet}}$ at $\mu_0 H$ = 0.05 T in the 10-nm-thick film. The ferromagnetic ordering temperature is estimated as $T_c \sim 200$ K.

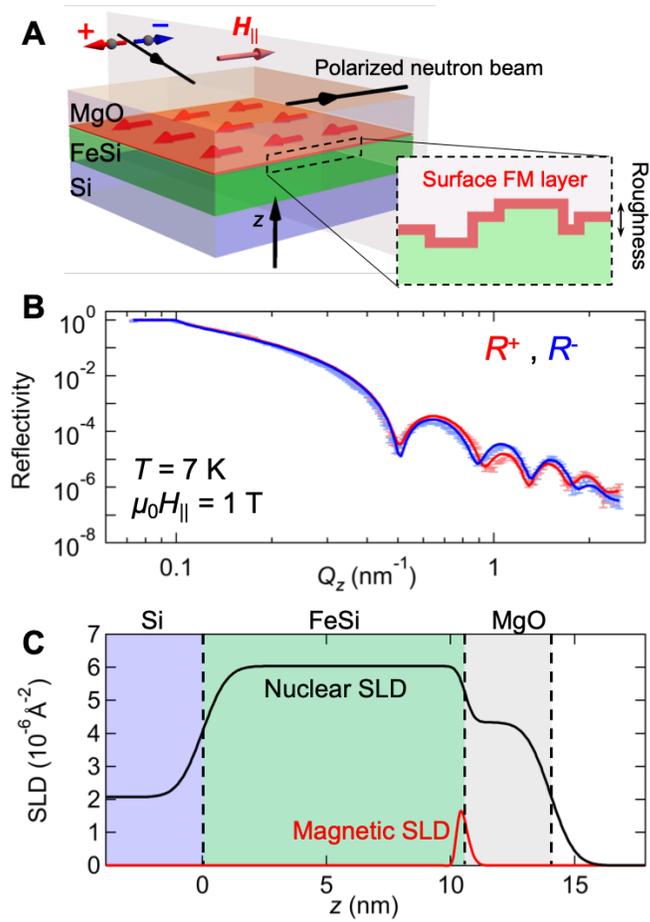

**Fig. 2. Observation of surface ferromagnetic state by polarized neutron reflectometry (PNR).** (**A**) Schematic illustrations of the experimental setup for PNR measurement and the distribution of ferromagnetic moments. The spins (red arrows) are aligned in the film plane by the in-plane magnetic field $H_\parallel$. (**B**) Measured (dots with error bars) and fitted (solid lines) reflectivities of polarized neutrons with spin-up and spin-down ($R^+$ and $R^-$) at $T = 7$ K and at $\mu_0 H_\parallel = 1$ T. (**C**) Depth profiles of the nuclear (black) and the magnetic (red) scattering length densities (SLDs).

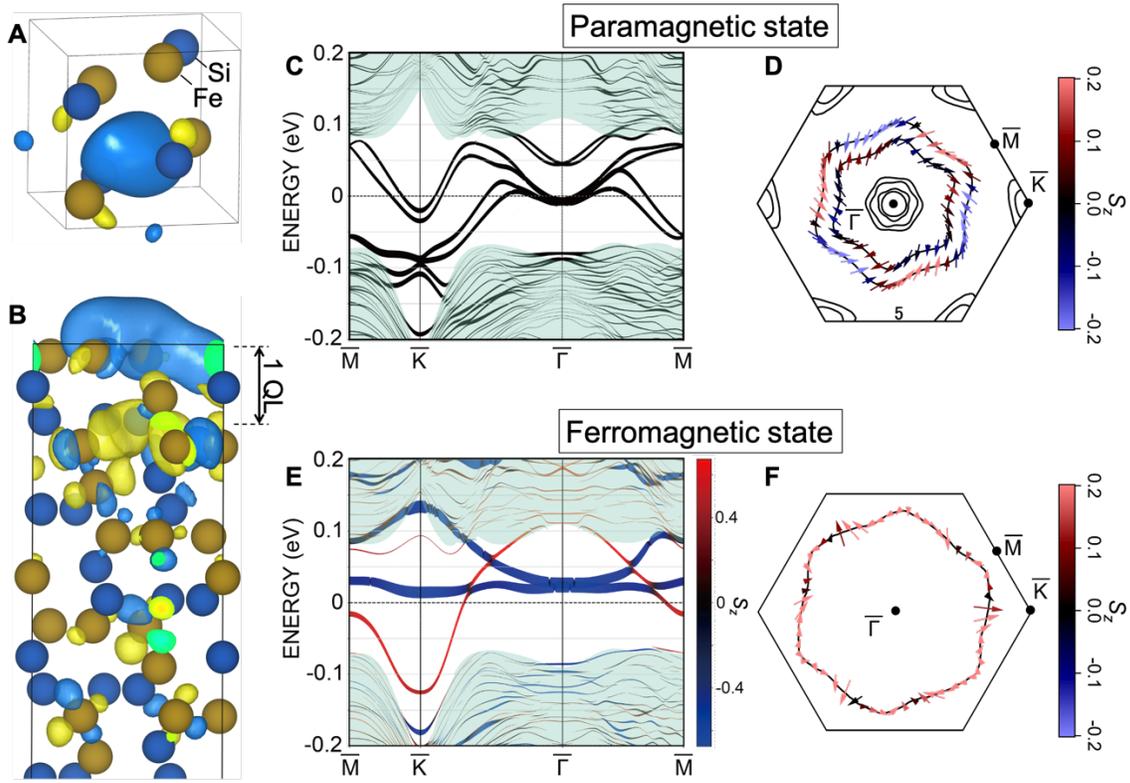

**Fig. 3. Band structures, spin-polarized Fermi contours and eigenfunction of the surface state.** (**A**) Wannier function for valence bands in the bulk FeSi. (**B**) Eigenfunction of the surface state of the paramagnetic FeSi slab with (111) plane. (**C**–**F**) Projected band structures and spin-polarized Fermi contours of the paramagnetic (C and D) and ferromagnetic (E and F) FeSi slab with SOC. In panels (C and E), the line thickness is proportional to the weight of the top and third Fe layers and the second and fourth Si layers. The green region corresponds to the bulk state. Some bands appear to be partially missing due to vanishingly little surface contribution. The energy is measured from the Fermi level. In panel (C), doubly-branched surface bands (thick curves) are separated approximately by 35 meV. In panel (E), color of the band represents the *z*-component of spin polarization. In panels (D and F), the direction and the color of arrows represent the *xy*- and the *z*-components of spin polarization, respectively.

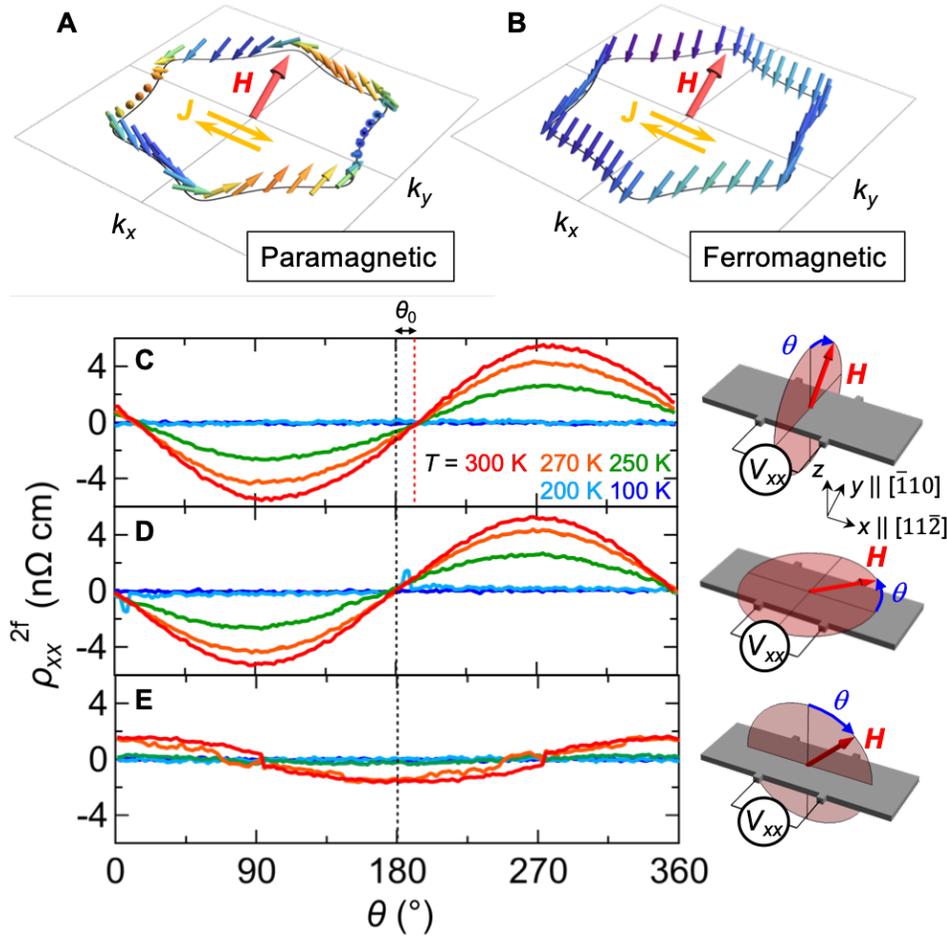

**Fig. 4. Detection of momentum-dependent spin polarization on Fermi contour by unidirectional magnetoresistance measurement in the 5-nm-thick film.** (**A** and **B**) Schematic illustrations of the unidirectional magnetoresistance measurements in the presence of the spin-polarized Fermi contours of the paramagnetic (A) and ferromagnetic (B) surface states. While the direction of the accumulated spins is almost reversed with the reversal of the current direction in the paramagnetic state (A), it remains nearly unchanged, pointing along the magnetic field $H$, in the ferromagnetic state (B). Note that spin angular momentum points opposite to magnetic moment. (**C**–**E**) Angular dependences of the second-harmonic resistivity $\rho_{xx}^{2f}$ for the scans in $yz$ plane (C), in $xy$ plane (D) and in $zx$ plane (E) at various temperatures. The magnitude of $H$ is fixed at $\mu_0 H = 1$ T and the $H$-angle is defined as illustrated in right sides of panels (C–E).

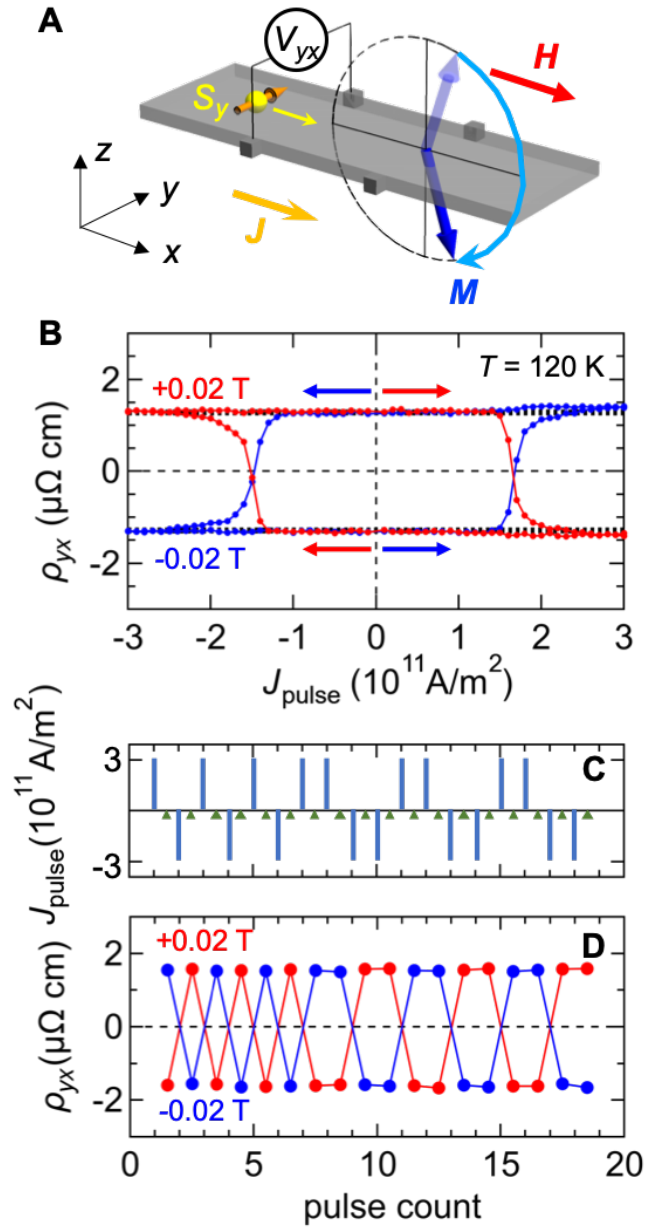

**Fig. 5. Deterministic magnetization switching by current pulses in the 5-nm-thick film.** (**A**) Schematic illustration of current-induced magnetization switching due to a dampinglike spin-orbit torque. (**B**) Current-pulse amplitude $J_{\text{pulse}}$ dependence of Hall resistivity $\rho_{yx}$ under small in-plane magnetic fields $\mu_0 H_x = \pm 0.02$ T at 120 K. The horizontal dashed lines represent the magnitude of anomalous Hall resistivity of the fully polarized state along $\pm z$. The switching ratio reaches almost 100%. (**C** and **D**) The deterministic magnetization switching by a series of current pulses. The procedure of current pulse injections (blue bars) and subsequent monitoring of the magnetization direction by anomalous Hall resistivity measurements (green triangles) is shown in panel (C). Hall resistivity proportional to the magnetization direction repeatedly switches with current pulse injections and the switching behavior is reversed with the reversal of in-plane magnetic field.

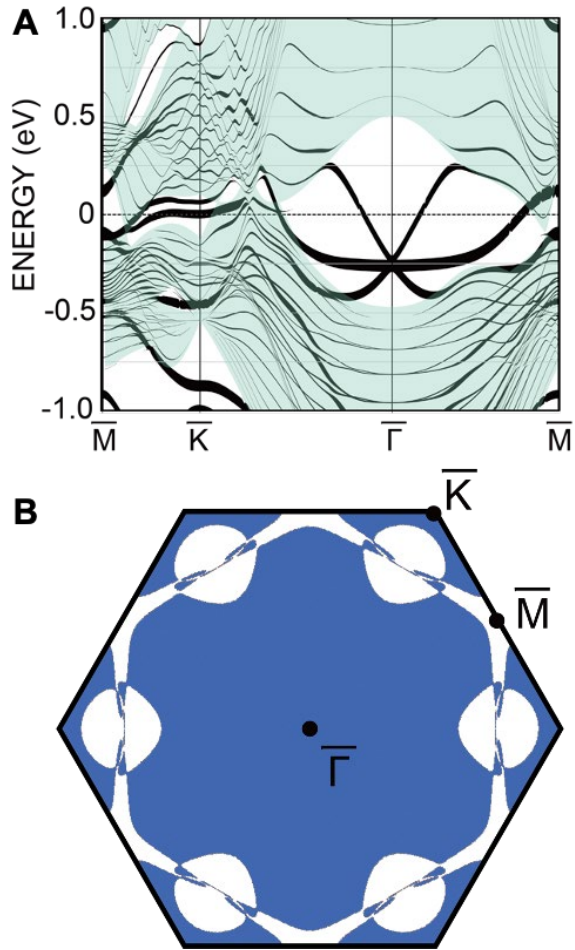

**Fig. 6. Relation between the surface band and the Zak phase in fcc FeSi.** (**A**) Projected band structure of a 15-layer slab of fcc FeSi with (111) plane. The calculations were performed on the paramagnetic state without SOC. The line thickness is proportional to the weight of the top and third Fe layers and the second and fourth Si layers. The green region corresponds to the bulk state. Some bands appear to be partially missing due to vanishingly little surface contribution. The energy is measured from the Fermi level. (**B**) Color map of the Zak phase in the surface Brillouin zone. The blue region represents the surface momenta $k_\parallel$ with the Zak phase $\theta(k_\parallel) = \pi$, while the other region is of the Zak phase $\theta(k_\parallel) = 0$.

**Supplementary Materials**

fig. S1. Characterization of FeSi thin films by x-ray diffraction (XRD), cross-sectional transmission electron microscopy (TEM) and energy dispersive x-ray spectroscopy (EDX).

fig. S2. Fidelity comparison of candidate models assuming different magnetization distributions for analyzing the polarized neutron reflectivities.

fig. S3. Theoretical calculation of depth dependence of magnetization.

fig. S4. Band structure of bulk FeSi and surface band structure for a different termination.

fig. S5. Cap-layer dependence of transport properties.

fig. S6. Absence of magnetization switching phenomena under magnetic fields perpendicular to current.

fig. S7. Thickness dependence of transport and magnetization properties in the as-grown FeSi thin films.

fig. S8. Switching of perpendicular magnetization by current pulses in the absence of external magnetic fields in the as-grown FeSi thin film.

fig. S9. Band structure of bulk FeSi with a fcc lattice structure.

fig. S10. Detailed temperature and magnetic-field dependence of Hall conductivity and magnetoresistance in the MgO-capped FeSi thin films with various thickness.

fig. S11. Magnetic-field, temperature and current density dependence of second harmonic resistivity in the MgO-capped FeSi thin film ($t = 5$ nm).

# Supplementary Materials for

# Emergence of spin-orbit coupled ferromagnetic surface state derived from Zak phase in a nonmagnetic insulator FeSi

Yusuke Ohtsuka, Naoya Kanazawa*, Motoaki Hirayama, Akira Matsui, Takuya Nomoto, Ryotaro Arita, Taro Nakajima, Takayasu Hanashima, Victor Ukleev, Hiroyuki Aoki, Masataka Mogi, Kohei Fujiwara, Atsushi Tsukazaki, Masakazu Ichikawa, Masashi Kawasaki, Yoshinori Tokura

*Corresponding author. Email: kanazawa@ap.t.u-tokyo.ac.jp

**This PDF file includes:**

Figs. S1 to S11

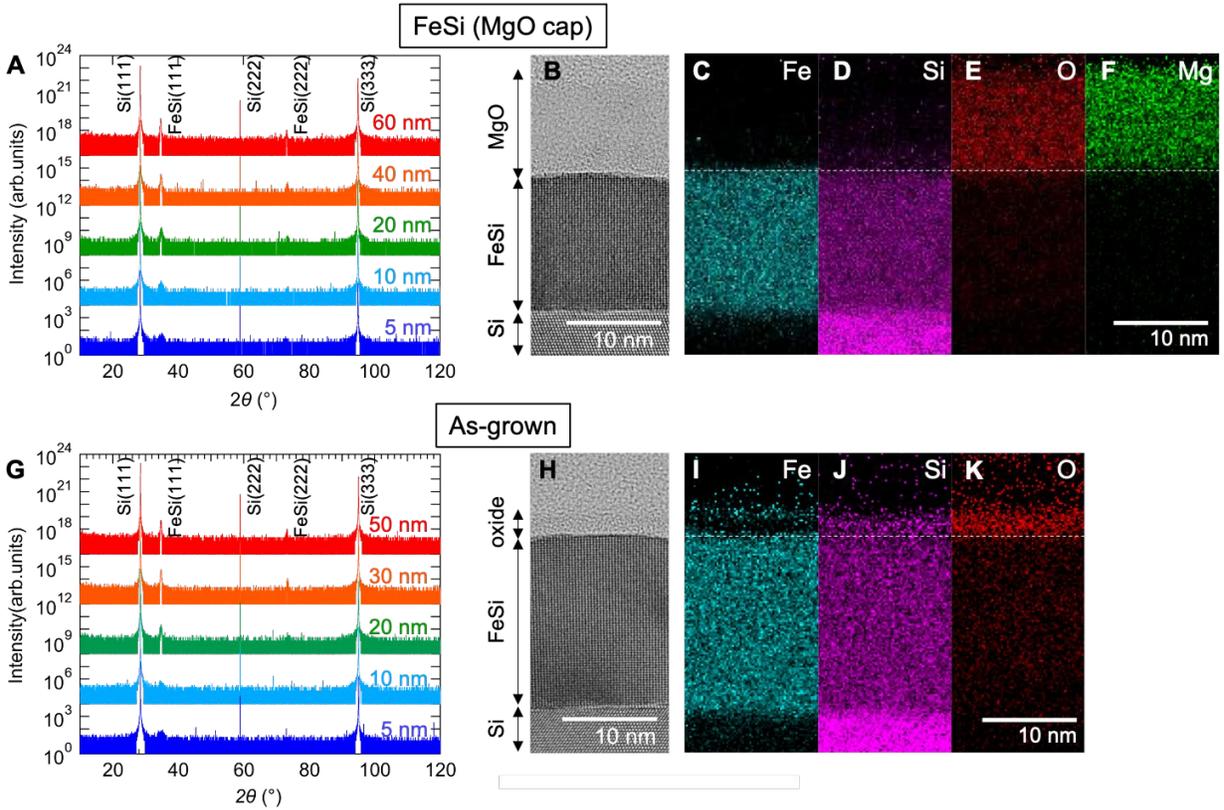

**Fig. S1.**
**Characterization of FeSi thin films by x-ray diffraction (XRD), cross-sectional transmission electron microscopy (TEM) and energy dispersive x-ray spectroscopy (EDX).** (**A-F**) Data set for the MgO-capped FeSi thin films, containing $\theta$-$2\theta$ XRD patterns (A), the TEM image (B) and EDX mappings for Fe (C), Si (D), O (E) and Mg (F) elements. Most of the MgO capping layer is amorphous. (**G-K**) Data set for the as-grown (uncapped) FeSi thin films, containing $\theta$-$2\theta$ XRD patterns (G), the TEM image (H) and EDX mappings for Fe (I), Si (J) and O (K) elements. While the sharp interface between FeSi and MgO is realized, the superficial layers of a few nanometers in the as-grown film are naturally oxidized by air exposure.

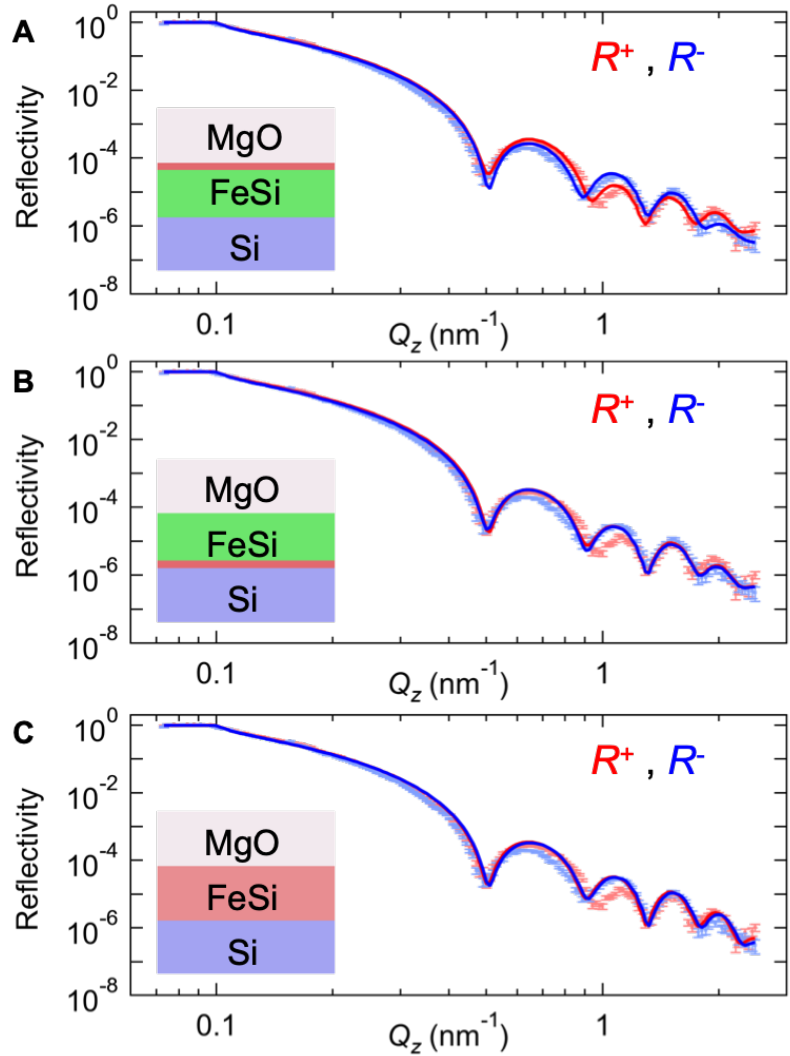

**Fig. S2.**

**Fidelity comparison of candidate models assuming different magnetization distributions for analyzing the polarized neutron reflectivities.** (A-C) The fitting analyses on the models with magnetization distributions at FeSi top layer (A), bottom layer (B), and in whole (C). The splitting between $R^+$ and $R^-$ data (light-colored dots with error bars) is well reproduced by the fitting curves (solid lines) in panel (A). By contrast, adding finite magnetization in the bottom layer (B) or whole the FeSi layer (C) does not reproduce the observed data, and therefore the possible magnitudes of the magnetization, which accounts for the splitting between $R^+$ and $R^-$, in models (B) and (C) were reduced to nearly zero by the least-squares fitting analysis. It is obvious that the model with surface magnetization on top of FeSi is the highest in fidelity.

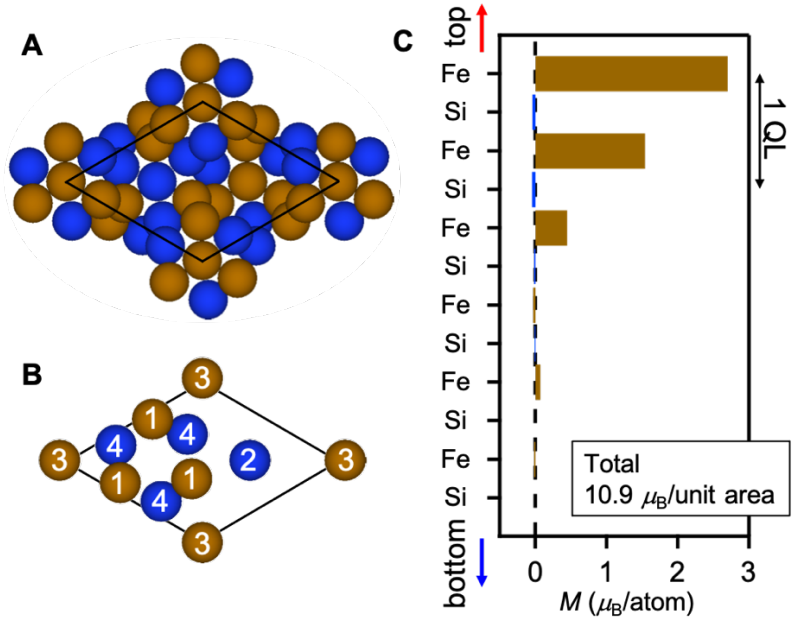

**Fig. S3.**
**Theoretical calculation of depth dependence of magnetization.** (**A**) Atomic arrangements in FeSi viewed from [111] direction. (**B**) Atomic arrangements in a 1 QL of FeSi. The label numbers represent the layer numbers counting from the top surface. In panels (A and B), brown and blue spheres represent Fe and Si atoms, respectively, and the diamond indicates the unit surface area. (**C**) Depth dependence of calculated magnetic moment $M$ per each atom. The total magnetization in the unit surface area yields 10.9 $\mu_B$. This value is in good agreement with the PNR measurements (11.7 $\mu_B$/unit area with an error of +4.1 $\mu_B$/unit area or −5.5 $\mu_B$/unit area).

On the other hand, the calculated $M$ is almost half of $M$ (~ 20 $\mu_B$/unit area) estimated by the magnetization measurements (Fig. 1F). This discrepancy may stem from that the magnetization on the side surfaces of bumps and holes inherent in FeSi thin film, which can be only detected by magnetization measurements.
4

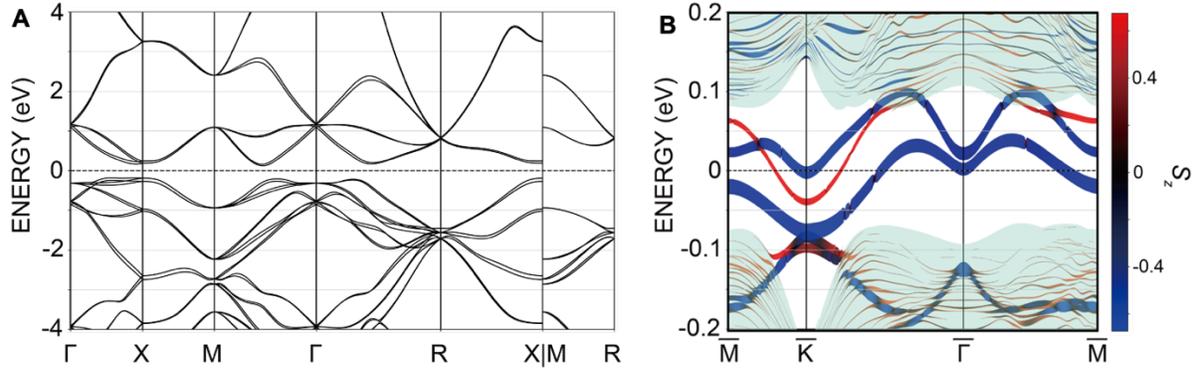

**Fig. S4.**
**Band structure of bulk FeSi and surface band structure for a different termination.** (**A**) Electronic band structure of bulk FeSi with the SOC. Since FeSi has broken inversion symmetry (space group: $P2_13$), FeSi has a spin-split band structure. (**B**) Projected band structure of ferromagnetic FeSi slab for a different termination with the SOC. The slab contains 15 repetitions of a stacking unit which is a quadruple layer (QL) consisting of Fe-sparse, Si-dense, Fe-dense and Si-sparse layers in the order from top to bottom. The line thickness is proportional to the weight of the top and third Fe layers and the second and fourth Si layers. The green region corresponds to the bulk state. Some bands appear to be partially missing due to vanishingly little surface contribution. The energy is measured from the Fermi level. The ferromagnetic-metal surface state also exists in this termination condition.

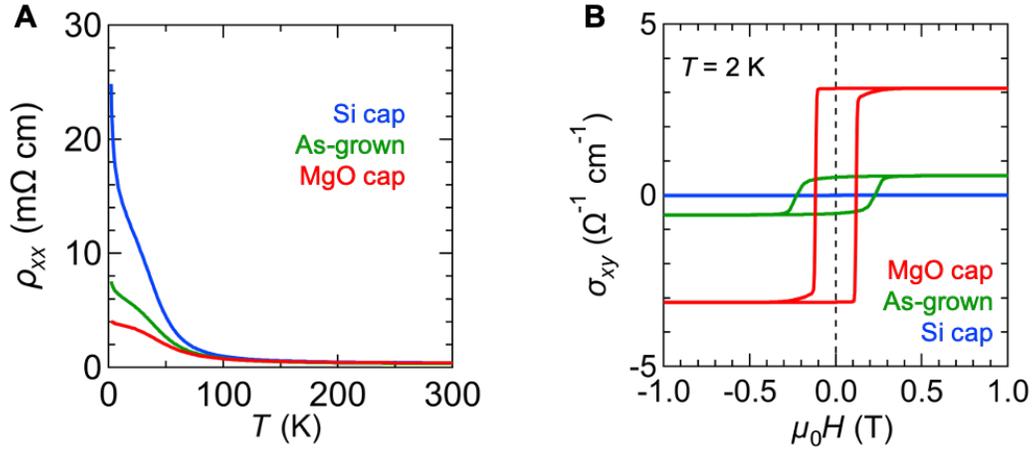

**Fig. S5.**
**Cap-layer dependence of transport properties.** (**A** and **B**) Temperature dependence of resistivity $\rho_{xx}$ (A) and magnetic-field dependence of Hall conductivity $\sigma_{xy}$ (B) in MgO-capped (red), Si-capped (blue) and as-grown (green) FeSi thin films. In every film, the nominal thickness of FeSi is 20 nm. The large magnitude of $\rho_{xx}$ and the absence of anomalous Hall effect indicate the disappearance of surface conductance and ferromagnetic order in the Si-capped film.

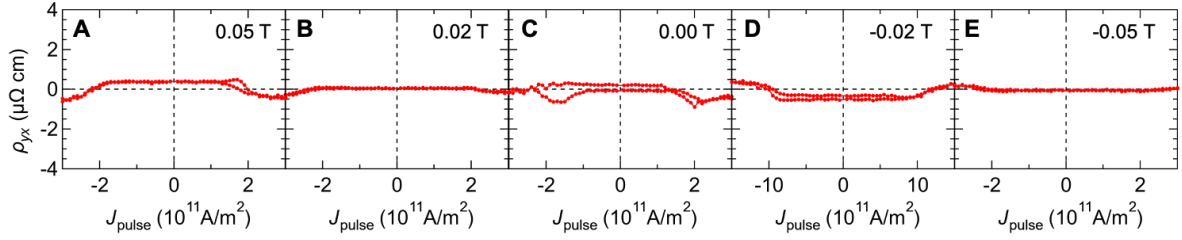

**Fig. S6.**
**Absence of magnetization switching phenomena under magnetic fields perpendicular to current.** (**A-E**) Magnetization switching was not realized by injecting current pulses with its magnitude below $3.0 \times 10^{11}$ A/m$^2$. This indicates the fieldlike spin-orbit torque (SOT) $\tau_{FL} = \boldsymbol{M}_z \times \boldsymbol{S}_y$ is not responsible for the magnetization switching.

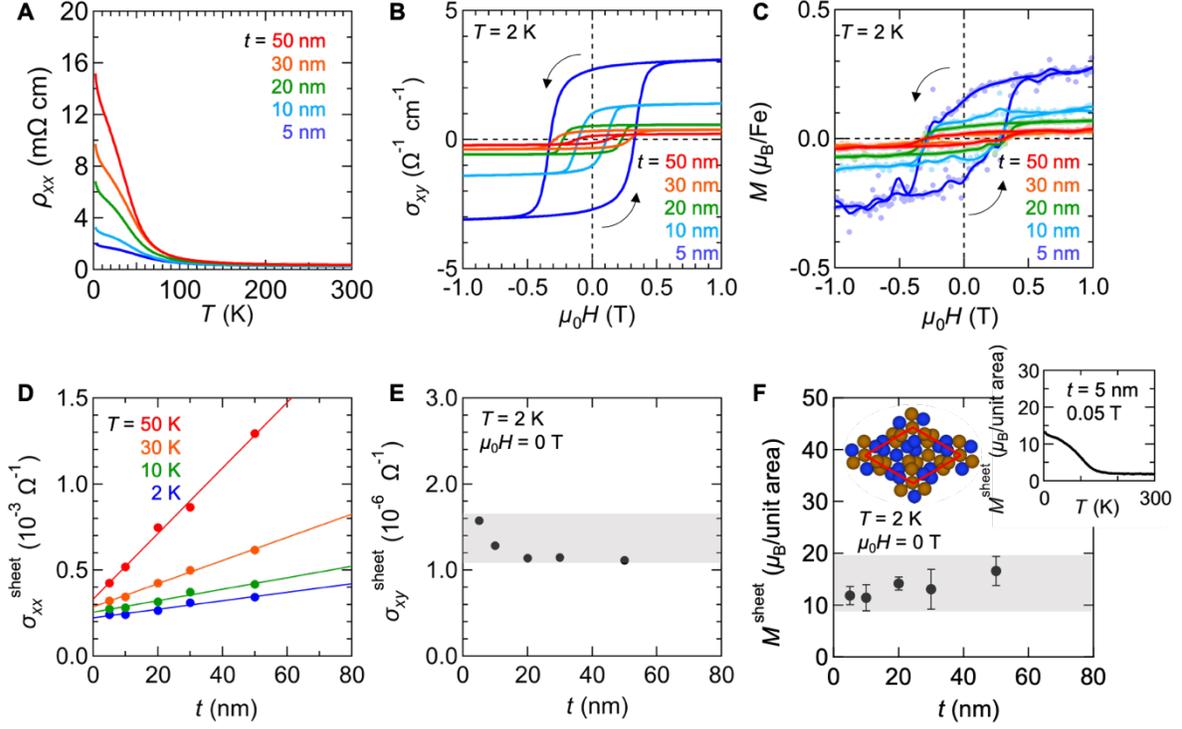

**Fig. S7.**
**Thickness $t$ dependence of transport and magnetization properties in the as-grown FeSi thin films.** (**A-C**) Temperature ($T$) dependence of resistivity $\rho_{xx}$ at zero magnetic field (A), and magnetic-field ($H$) dependences of Hall conductivity $\sigma_{xy}$ (B) and magnetization $M$ (C) at $T = 2$ K in the as-grown FeSi thin films with various $t$. Here, the physical quantities per volume ($\rho_{xx}$, $\sigma_{xy}$ and $M$) are estimated under an assumption of the uniform electrical conductance and magnetization across the samples. The large $t$-dependences of their magnitudes indicate the heterogeneity of transport and magnetization properties. (**D-F**) Thickness dependences of sheet conductance $\sigma_{xx}^{\text{sheet}}$ (D), sheet anomalous Hall conductance $\sigma_{xy}^{\text{sheet}}$ (E) and remanent magnetization per surface unit cell area $M^{\text{sheet}}$ (F) at various $T$. Color solid lines or thick gray lines in each panel are guides to the eye. The enhanced $\sigma_{xy}^{\text{sheet}}$ in the thinner films may be attributed to the inaccurate evaluation of FeSi thickness due to the presence of oxidized surface layer. (See fig. S1.) The insets of panel (F) show the surface unit cell and $T$-dependence of $M^{\text{sheet}}$ at $\mu_0 H = 0.05$ T in the 5-nm-thick film. The ferromagnetic ordering temperature is estimated as $T_c \sim 120$ K.

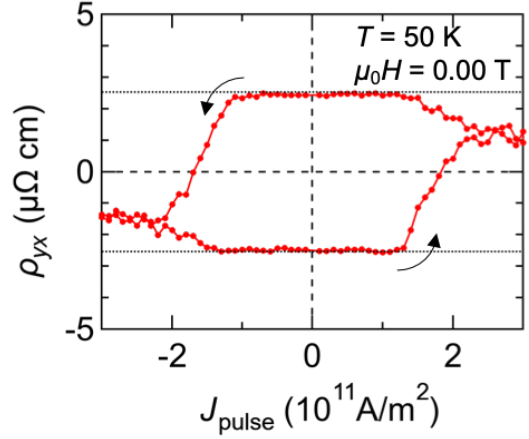

**Fig. S8.**
**Switching of perpendicular magnetization by current pulses in the absence of external magnetic fields in the as-grown FeSi thin film.** Current-pulse amplitude $J_{pulse}$ dependence of Hall resistivity $\rho_{yx}$ without external magnetic fields at $T = 50$ K. The horizontal dashed lines represent the magnitude of anomalous Hall resistivity of the fully polarized state along $\pm z$. The switching ratio is approximately 60 %. The threefold symmetric out-of-plane spin texture being reflective of (111)-surface geometry may serve as the effective spin-orbit (SO) field, resulting in the bias-field-free magnetization switching. As seen in fig. S5B, the out-of-plane magnetic anisotropy seems reduced in the as-grown film, compared to that of the MgO-capped film. Due to the reduced perpendicular anisotropy, the $z$-component SO field may play a critical role in the magnetization switching. Note that the direction of SO field is switched between $+z$ and $-z$ by the reversal of current direction, which also enables the deterministic switching.

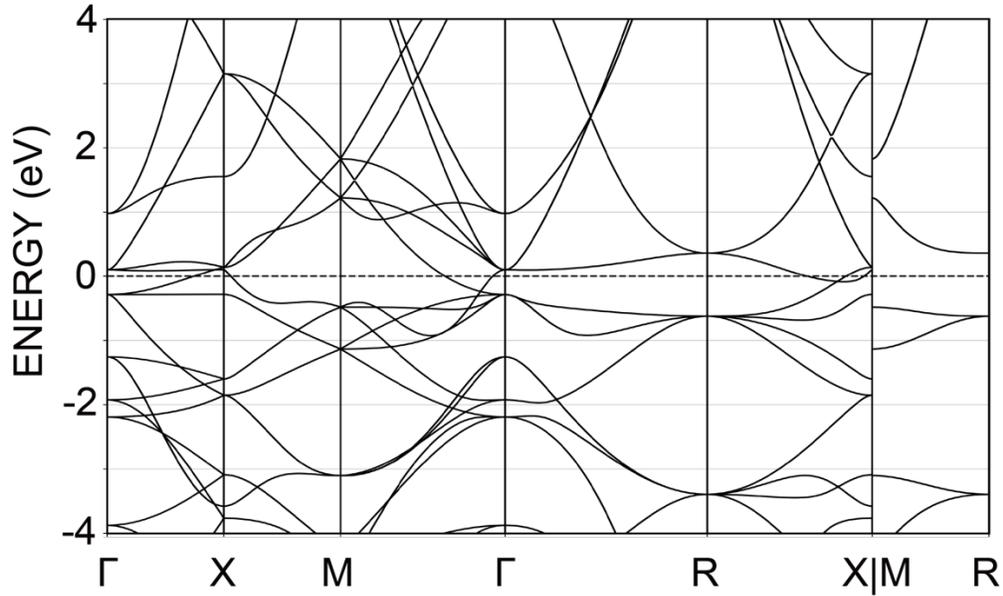

**Fig. S9.**
**Band structure of bulk FeSi with a fcc lattice structure.** Here, for comparison, we use the same size of unit cell as *B*20, instead of the primitive cell of the fcc structure. The SOC is not included. The energy is measured from the Fermi level. Most of the regions, such as around the M and R points, remain gapped without band inversion during the change from the original *B*20 structure to the fcc structure.

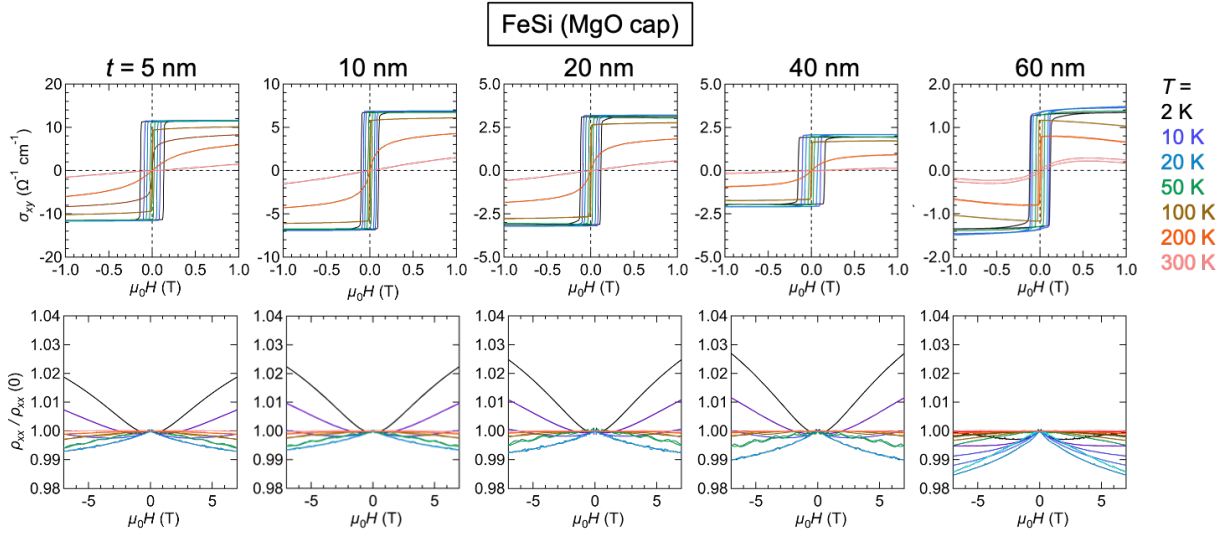

**Fig. S10.**

**Detailed temperature and magnetic-field dependence of Hall conductivity and magnetoresistance in the MgO-capped FeSi thin films with various thickness.**

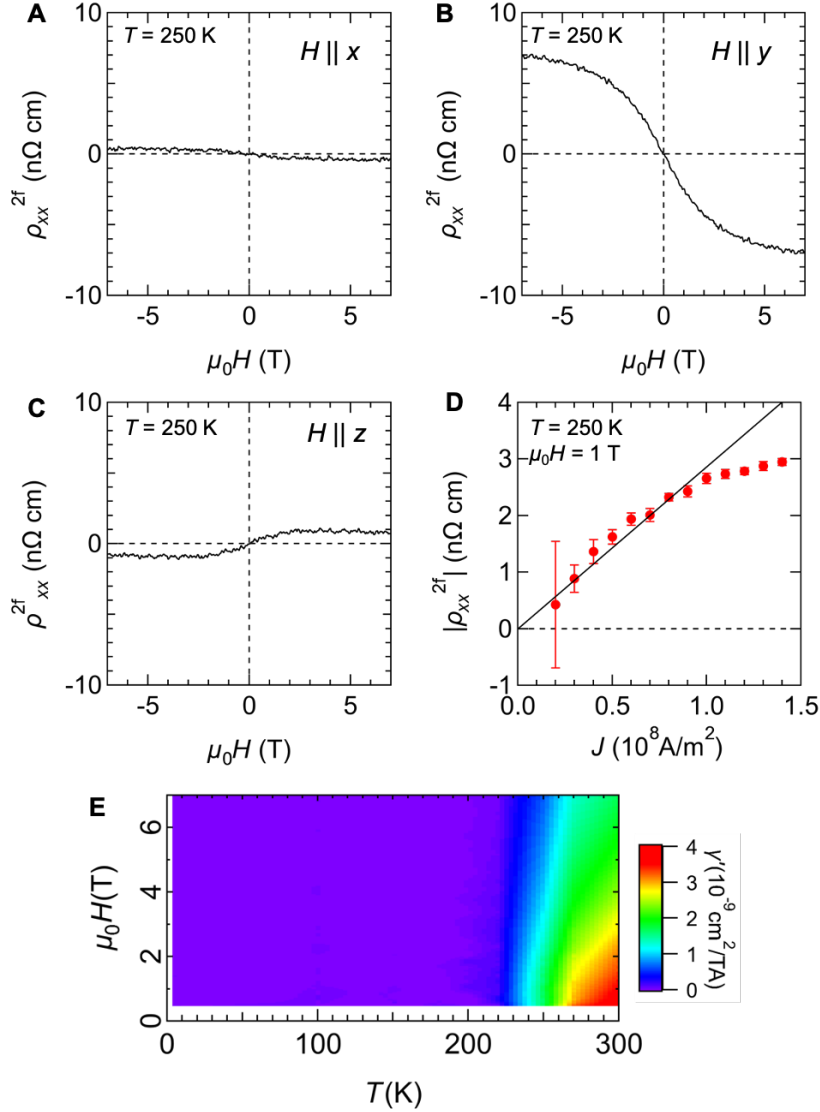

**Fig. S11.**
**Magnetic-field ($H$), temperature ($T$) and current density ($J$) dependence of second harmonic resistivity $\rho_{xx}^{2f}$ in the MgO-capped FeSi thin film ($t$ = 5 nm).** (A-C) Magnetic-field dependence of $\rho_{xx}^{2f}$ under $H \parallel x$ (A), $H \parallel y$ (B) and $H \parallel z$ (C). (D) Current-density dependence of $\rho_{xx}^{2f}$ under $\mu_0 H$ = 1 T along $y$-direction. The coordinate bases are defined in Fig. 3 of the main text. Deviation from the linear dependency on $H$ and $J$ is observed above $\mu_0 H$ = 2 T and above $J$ = 1.0 × 10$^8$ A/m². (E) Color map of the coefficient $\gamma' = -2\rho_{xx}^{2f}/\rho_0 HJ$ under $H \parallel y$. The unidirectional magnetoresistance is present above the ferromagnetic ordering temperature $T_c$ ~ 200 K.

12